\documentclass[envcountsame,envcountsect,runningheads]{llncs}
\usepackage[T1]{fontenc}
\usepackage[english]{babel}
\usepackage{algorithm}
\usepackage{algorithmic}

\usepackage{amsmath,amsfonts,amssymb}
\usepackage{graphics, graphicx}
\usepackage{calc}

\usepackage{xurl,hyperref}
\usepackage[table]{xcolor}
\usepackage{multirow}
\usepackage{comment}
\usepackage{listings}
\usepackage{graphicx}
\usepackage{color}
\usepackage{tikz} 

\title{Community Covert Communication -  \\ Dynamic Mass Covert Communication\\ Through Social Media\thanks{This work has been presented at 44CON 2024 \& 44CON 2025 in London.}}  
\author{Eric Filiol}
\institute{Retired Professor - Independent Researcher, Paris, France\\  \email{efll@protonmail.com} \\ \url{https://ericfiliol.site}}
\date{\today} 

\begin{document}
\maketitle

\begin{abstract}
Since the early 2010s, social network-based influence technologies have grown almost exponentially.
Initiated by the U.S. Army's early OEV system in 2011, a number of companies specializing in this field have emerged. 
The most (in)famous cases are Bell Pottinger, Cambridge Analytica, Aggregate-IQ and, more recently, Team Jorge. 

In this paper, we consider the use-case of sock puppet master activities, which consist in creating hundreds or even thousands of avatars, 
in organizing them into communities and implement influence operations. On-purpose software is used to automate these operations (e.g. Ripon software, AIMS) and organize these avatar populations into communities. The aim is to organize targeted and directed influence communication to rather large communities (influence targets).

The goal of the present research work is to show how these community management techniques (social networks) can also be used to communicate/disseminate relatively large volumes (up to a few tens of Mb) of multi-level encrypted information to a limited number of actors.
To a certain extent, this can be compared to a Dark Post-type function, with a number of much more powerful potentialities. As a consequence, the concept of communication has been totally redefined and disrupted, so that eavesdropping, interception and jamming operations no longer make sense.
\end{abstract}

\section{Introduction}
When transmitting data in a secure way (providing confidentiality AND hiding the transmission channel itself) different technologies are possible, all of them described by covert channel communication~\cite{filiol_bsides}. The most widely known technique among many others is steganography~\cite{fridrich}. 

According to NATO terminology, two different aspects must be taken into account in Information and Communication Security:
\begin{itemize}
	\item COMSEC (\textit{Communication Security}) ensures the security (confidentiality and integrity) of telecommunications. COMSEC refers to the security of information transmitted regardless of the communication channel. It includes different technical and scientific domains such as cryptography, Tempest/EMSEC, physical security (network and rooms).
	\item TRANSEC (\textit{Transmission Security}) ensures the protection of the communication channel itself and especially the existence of secret data being exchanged (prevent interception/eavesdropping/wiretapping, disruption of reception, jamming, communications deception, and/or derivation of intelligence by analysis of transmission characteristics such as signal parameters or message externals...).
	TRANSEC hence deals with the security of communication part (the channel), rather than that of the information being communicated. To enforce it, common techniques are steganography/covert channels, traffic flow security, routing protocols and many others.
\end{itemize}

However, in any communication, according to the classic model~\cite{shannon48}, the TRANSEC part suffers from two fundamental weaknesses on which any attack is based (regardless of its effectiveness).
\begin{itemize}
	\item The existence of a channel. Transmitting information by nature is a weakness.
	\item The unicity of the channel. Some attempts, such as frequency hopping spread spectrum, intend to significantly reduce this problem, by the alternative use of several channels (sub-carriers) distributed across a frequency band according to a pseudo-random sequence known to the transmitter and receiver. But, to a certain extent, transmitting radio signals by rapidly and alternatively changing the carrier frequency among many frequencies occupying a large spectral band, can be reduced to the case of a single channel (divided into a series of sub-channels). Some attacks have shown that this technique can be defeated \cite{fhattack2,fhattack1}.
\end{itemize}
In this paper, we consider a completely different approach in which the channel and its uniqueness disappear as such. The information, previously encrypted using the deniable encryption technique presented in \cite{filiol44con,filiol_bsides}, is no longer accessible via a single channel as defined by Shannon. To achieve this, we use social networks and the concept of community. The encrypted information is encoded by interactions between individuals, which can be dynamically reconfigured by an actor with the ability to create and manage these communities on existing social networks in real time. To this end, we have considered the techniques of influence on social networks used by some companies around the world to influence electoral processes (sock puppet techniques) and adapted them to our new TRANSEC vision. 

As a result, our approach called \textit{Community Covert Communication} (CCC) enables to give access to a unique encrypted content $C$ of several tens of Mb to $\alpha$ users, each of them getting access to a unique plaintext $P_i$ with $1 \leq i \leq \alpha$ thanks to 
a unique, deterministic encryption algorithm $E$. Each user has a secret key $K_i$ ($1 \leq i \leq \alpha$) and each plaintext $P_i = E(K_i, C)$ has the same size as the ciphertext. In other words $|C| = |P_i|$ ($1 \leq i \leq \alpha$). The content $C$ can be dynamically changed by reconfiguring the encoding support community.

This paper is organized as follows. In Section~\ref{s1} we present the concept and state-of-the-art regarding sock puppet master activities and how social networks have been or are currently used for influence operations. In Section~\ref{s2}, we give a concise presentation of what graphs are as mathematical objects, how to code them in practice and show how graphs are used to represent and manipulate social networks and communities. This section also briefly presents what hypergraphs are and to consider more complex organisations of social networks communities. In Section~\ref{s3} we present the \textit{Community Covert Communication Technique} (CCCT) allowing multi-level encrypted communications towards a limited number of actors using non trivial deniable encryption. Section~\ref{s4} addresses the security of CCCT. Finally in Section~\ref{conc} we summarize our results and present future work.

\textbf{Disclaimer.-} \textit{Any views, opinions and materials presented in this paper are personal and are the results of the author's own research work. They belong solely to the author and, in any case, they do not represent those of people, institutions, companies or organizations that the author may or may not be associated with in professional or personal capacity (including past, present and future employers).} 
\section{Context and Use-Case of Sock Puppet Masters} \label{s1}
\subsection{History and Current Situation}
Influence operations are as old as the history of the world, but the advancement of communication technologies and then digital communications, networks, the internet, and AI have enabled unparalleled scalability.

Since operations such as Operation Mockingbird by the CIA in the late 1960s, involving the print media \cite{mockingbird1,mockingbird2}, which was the prehistory of such operations, many others have been regularly identified and revealed, each illustrating progress in terms of volume, technical sophistication, and automation. The number of projects revealed is becoming particularly significant of an intensifying trend. A (non-exhaustive) list is maintained on \cite{stateinfluenceops}, but some operations are missing (in or by South Africa, India, Israel, Russia, for example).

Certain operations caught our attention because, on the one hand, they use social media on an unprecedented scale and, on the other hand, they rely on complex but powerful software.
\begin{itemize}
\item \textbf{Operation Earnest Voice (OEV)}. This is a psychological operation conducted by the US Central Command (CENTCOM) that uses fake accounts (\textit{personae} or \textit{sock puppet} account) to spread pro-American propaganda on targeted social media networks based outside the United States. In 2011, the US government signed a \$~2.8 million contract with web security company \textit{Ntrepid} to develop specialized software enabling US government agents to publish propaganda on ``foreign language websites using fake accounts (sock puppet accounts)''~\cite{centcom,centcom2}.
\item \textbf{\#ChinaAngVirus disinformation campaign} (CIA, 2019). The CIA allegedly used social media accounts with fake identities to spread rumors about the Chinese government, according to an investigation. in fact, this has been revealed by a Reuters report as a covert Internet anti-vaccination propaganda and disinformation campaign conducted by the United States Department 
of Defense at the height of the COVID-19 pandemic from the spring of 2020 to the spring of 2021, to dissuade Filipino, Central Asian, and Middle Eastern citizens from receiving Sinovac Biotech's CoronaVac vaccine and from using other Chinese COVID-19 medical supplies. The propaganda campaign used at least 300 fake accounts on Twitter, Facebook, Instagram, and other social media websites meant to look like local internet users \cite{reuters1,reuters2,wikipedia2}. 
\item \textbf{Facebook - Cambridge Analytica scandal} and the Cambridge Analytica network (SCL, AggregateIQ, Team Jorge and others). Since 2010, influence techniques have been increasingly used to modify or influence, often successfully, the results of national elections in a large number of countries (at least 50 major national elections have been identified to date, including the Brexit, the 2016 US elections, and the Nigerian elections). The companies involved have often changed their names, branched out, and others have emerged, but the development of these technologies, particularly by Israeli companies specializing in them, continues. It is attracting more and more political parties\footnote{For instance, during the French presidential elections and then the legislative elections in 2017 and 2022, all political parties used such services or software, either off-the-shelf (such as Nation Builder) or custom-made \cite{france-info}}.
\end{itemize}
\subsection{Technical Aspects and known Features}
The basic tool for these operations is to have a number of fake accounts corresponding to fake profiles. Such accounts are called \textit{sock puppets} or \textit{sock puppet accounts}. They correspond to false online identities used for deceptive purposes. Sock puppets include online identities created to praise, defend, or support a person or an organization, to manipulate public opinion or to circumvent restrictions such as viewing a social media account that a user is blocked from. Sock puppets are unwelcome in many online communities and forums \cite{wikisockpuppet}. 

\textit{Sock puppet masters} (or puppeteers) are people that can generate, manage and animate armies of sock puppet accounts, in a dynamic and adaptative way, thanks to specialized software or APIs. 

The general principle is to be able to generate and manage an army of fake profiles (sock puppets) in real time, to carry out micro-targeting (dark post function), supported by artificial intelligence to write viral posts on demand and circulate them on Facebook, but also on Twitter, LinkedIn, Telegram, etc. In other words, the goal is to manage and synchronize several different communities on different social media platforms \textit{via} a single API. This brings together several social networks into a single meta-network where each social network appears as a subset of communities. 

It is essential to bear in mind that the content of profiles and their communications (posts) are just as important as the connections between profiles (the relationships they maintain). It is the density of these connections within a given group that defines the concept of \textit{community}. The CCC technique presented in Section~\ref{s3} is based precisely on the ability to think in terms of communities.

For these operations, specialized software for the dynamic generation and management of fake accounts (or \textit{personae}) has been developed and has been used on a large scale for several years (and still are). The main cases identified (and their known features) are:
\begin{itemize}
	\item \textbf{Operation OEV and NTrepid software}. Fifty ``operator'' user licenses, 10 fake accounts controllable by each user. Other features are available on \cite{centcom3}.
	\item \textbf{Facebook - Cambridge Analytica}. The core software, Ripon~\cite{UKreport2019,ripon}, is a software platform incorporating artificial intelligence, integrating algorithms that use stolen personal data from Facebook accounts to create psychographic profiles used for micro-targeted election campaigns aimed at influencing voters' emotions in order to change their voting intentions or encourage them to abstain from voting. Various factors and information suggest that Ripon is capable of managing several thousands of accounts (around 3,000), but it is likely that this activity has enabled Tal Hanan of Team Jorge to design the AIMS software thanks to early collaboration with Cambridge Analytica.
	
	\item \textbf{Election manipulation and Team Jorge activities}~\cite{aims,teamjorge,teamjorgewiki}. AIMS (\textit{Advanced Impact Media Solutions)} software is currently considered as the most sophisticated known sock puppet master platform. Team Jorge is the name given to a team of subcontractors specializing in the use of cybermalicious activities, including hacking, sabotage, and disinformation campaigns on social media carried out by bot farms in order to manipulate election results. This software automates the centralized control of thousands of fake profiles on social media platforms such as Twitter, LinkedIn, Facebook, Telegram, TikTok, and others. A number of these profiles also have Amazon and Airbnb accounts, as well as credit cards and Bitcoin wallets. According to public data, AIMS is able of generating, managing, and animating communities of nearly 30,000 personae. Non-public information suggests sizes of up to 100,000 personae.
\end{itemize}
It is more than likely that new developments have pushed the capabilities of such software, platforms or API toward unprecedented performances. Nowadays, sock puppet mastery activities are largely developed for OSINT and SOCMINT activities (both passive and active\footnote{For instance Maltego offers SOCMINT capabilities (see 
\url{https://www.maltego.com/blog/ how-to-use-sock-puppet-accounts-to-gather-social-media-intelligence/}) or other similar services are proposing similar capabilities such as \url{https://www.sockpuppet.io/the-platform/}}, but with the fear that this use is increasingly being diverted to more questionable activities).

For the remainder of this article, and in light of all of the above, we will assume that managing communities of size $N = 50,000$ on social networks \textit{via} an API or sock puppet mastery software is now a common and standard practice.

\section{Graphs, Graph Coding Techniques and Social Networks} \label{s2}
\subsection{Basic Concepts of Graphs}
We assume that the reader has the necessary basic knowledge of graph theory and hypergraph theory. We will only review the concepts that are essential for the rest of the article. The reader can refer to \cite{berge1970,hgraph,gross2014} for a complete presentation of graph theory and hypergraph theory.
\begin{figure}
	\centering
	\scalebox{.9}{
		\begin{tabular}{cc}
			\begin{tikzpicture}[node distance={15mm}, thick, main/.style = {draw, circle},scale=.8]
				
				\node[main] (1) {$1$}; 
				\node[main] (2) [above right of=1] {$2$}; 
				\node[main] (3) [below right of=1] {$3$}; 
				\node[main] (4) [above right of=3] {$4$}; 
				\node[main] (5) [above right of=4] {$5$}; 
				\node[main] (6) [below right of=4] {$6$}; 
				\draw[->] (1) -- (2); 
				\draw[->] (1) -- (3); 
				\draw[->] (1) to [out=135,in=90,looseness=1.5] (5); 
				\draw[->] (2) -- (4); 
				\draw[->] (4) -- (3); 
				\draw[->] (5) -- (4); 
				\draw[->] (5) to [out=315, in=315, looseness=2.5] (3); 
				\draw[->] (6) -- node[midway, above right, sloped, pos=1] {} (4); 
			\end{tikzpicture} &
			\begin{tikzpicture}[node distance={15mm}, thick, main/.style = {draw, circle}, scale= 0.9] 
				\node[main] (1) {$1$}; 
				\node[main] (2) [above right of=1] {$2$}; 
				\node[main] (3) [below right of=1] {$3$}; 
				\node[main] (4) [above right of=3] {$4$}; 
				\node[main] (5) [above right of=4] {$5$}; 
				\node[main] (6) [below right of=4] {$6$}; 
				\draw (1) -- (2); 
				\draw (1) -- (3); 
				\draw (1) to [out=135,in=90,looseness=1.5] (5); 
				\draw (2) -- (4); 
				\draw (4) -- (3); 
				\draw (5) -- (4); 
				\draw (5) to [out=315, in=315, looseness=2.5] (3); 
				\draw (6) -- node[midway, above right, sloped, pos=1] {} (4); 
			\end{tikzpicture}
	\end{tabular}}
	\caption{Directed graph (left) and undirected graph $G$ (right)} \label{figgraph1}
\end{figure}
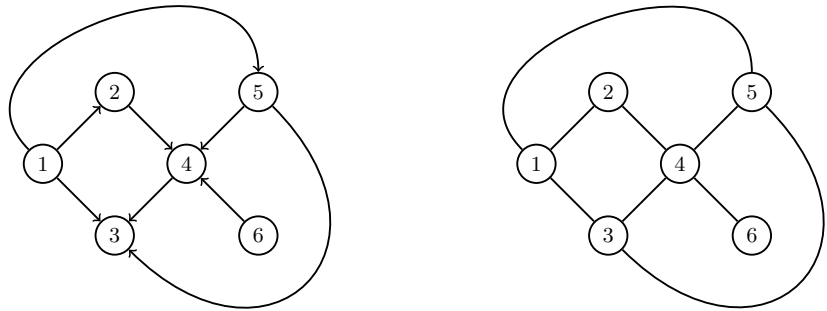

\begin{definition}
A (undirected) graph $G = (V, E)$ consists of two sets $V$ and $E$:
\begin{itemize}
	\item The elements of $V$ are called vertices (or nodes). 
	\item The elements of $E$ are called edges.
	\item Each $e \in E$ has a set of one (loop) or two (edge) vertices associated to it. 
\end{itemize}
\end{definition}
So $E \subsetneq \mathcal{P}(V)$. The order or cardinality of $G$ is given by $n = |V|$. The density of $G$ is given by $D = \frac{|V|}{\frac{|V|.(|V| - 1)}{2}}$
\begin{definition}
	A directed graph or digraph $\overrightarrow{G} = (V, A)$ consists of two sets $V$ and $A$:
	\begin{itemize}
		\item The elements of $V$ are called vertices (or nodes). 
		\item The elements of $A$ are called arcs.
		\item Each $\overrightarrow{a} \in A$ has a couple of one (directed loop) or two (arc) vertices associated to it.
	\end{itemize}
\end{definition}
So $A \subsetneq V\times V$. The order or cardinality of $\overrightarrow{G}$ is given by $n = |V|$. The density of $\overrightarrow{G}$ is given by $D = \frac{|V|}{|V|^2}$.
Vertices connected by an edge or an arc are said adjacent. Let us note that the density $D$, in both cases, express the ratio of adjacent vertices compared to the maximum possible edges or arcs. 
Figure~\label{figgraph1} shows a simple example of a digraph $\overrightarrow{G} = (V, A)$ and of simple graph $G = (V, E)$ with $V = \{1, 2, 3, 4, 5, 6\}$, $A = \{(1, 2), (1, 3), (1, 5), (2, 4), (4, 3), (5, 3), (5, 4), (6, 4)\}$ and $E = \{\{1, 2\}, \{1, 3\}, \{1, 5\}, \{2, 4\}, \{3, 4\}, \{3, 5\}, \{4, 5\}, \{4, 6\}\}$. We have $D(\overrightarrow{G}) = 0.22$ and $D(G) = 0.53$.
\begin{definition}
	A partial sub-graph of $G = (V, E)$ (respectively of $\overrightarrow{G} = (V, A)$) is a graph $G' = (V', E')$ ($\overrightarrow{G'} = (V', A')$) such that $V' \subseteq V$ and $E' \subseteq E$ (respectively $A' \subseteq A$).
\end{definition}
From that, we can define a community as a partial sub-graph $G'$ (resp. $\overrightarrow{G'})$ such that $D(G') > D(G)$ (resp. $D(\overrightarrow{G'}) > D(\overrightarrow{G})$).
\begin{figure}
	\centering
	\scalebox{.9}{
		\begin{tabular}{ccc}
			\begin{tikzpicture}[node distance={15mm}, thick, main/.style = {draw, circle},scale=.8]
				
				\node[main] (1) {$1$}; 
				\node[main] (2) [above right of=1] {$2$}; 
				\node[main] (3) [below right of=1] {$3$}; 
				\node[main] (4) [above right of=3] {$4$}; 
				\draw[->] (1) -- (2); 
				\draw[->] (1) -- (3);  
				\draw[->] (2) -- (4); 
				\draw[->] (4) -- (3); 
			\end{tikzpicture} &\hspace{0.5cm} &
			\begin{tikzpicture}[node distance={15mm}, thick, main/.style = {draw, circle}, scale= 0.9] 
				\node[main] (1) {$1$}; 
				\node[main] (2) [above right of=1] {$2$}; 
				\node[main] (3) [below right of=1] {$3$}; 
				\node[main] (4) [above right of=3] {$4$}; 
				\draw (1) -- (2); 
				\draw (1) -- (3);  
				\draw (2) -- (4); 
				\draw (4) -- (3); 
			\end{tikzpicture}
	\end{tabular}}
	\caption{Partial sub-graphs (directed, left - undirected, right)} \label{figgraph2}
\end{figure}
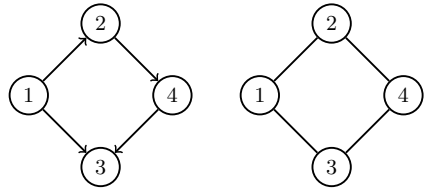
Figure~\label{figgraph2} shows partial subgraphs drawn from Figure~\ref{figgraph1} with $V' = \{1, 2, 3, 4\}$, $A = \{(1, 2), (1, 3), (2, 4), (4, 3), \}$ and $E' = \{\{1, 2\}, \{1, 3\}, \{2, 4\}, \{3, 4\}\}$. We have $D(\overrightarrow{G'}) = 0.25$ and $D(G) = 0.66$.
\subsection{Coding and Visualizing Social Networks with Graphs}
\subsubsection{Social Networks as Graphs}
In this section we dig into the modelling aspects of social networks. Social networks are essentially made up of three main components:
\begin{itemize}
	\item Individuals with one or more defining properties, characteristics or data. This data is contained in the profile and consists of information entered by individuals and metadata, usually added, generated by the system underlying and managing the social network.
	\item More or less complex, one-way or two-way relationships or interactions between individuals. These interactions may themselves have properties or contain information about the nature of these interactions or relationships. Again, these are either explicit data or metadata.
	\item \textit{Communities} of individuals. A community is a subset of individuals developing particular interactions and/or relationships centred on a subset of characteristics shared by these individuals among themselves rather than with all individuals in the social network.
\end{itemize}
The modelling, manipulation and visualization of social networks is traditionally done using graph theory \cite{furht2010} and, more recently, hypergraphs \cite{liu2010}. Table~\ref{tab1} summarises the correspondence between graphs and social networks.
\begin{table}[h]
\centering
\begin{tabular}{|c|c|} \hline
	Graph   & Social Network \\ \hline \hline
	Vertex  & Individual \\
	Edge/arcs & Relationship, interconnection \\
	Partial subgraph & Community \\ \hline
\end{tabular}
\caption{Correspondence between graphs and social networks (simplified)} \label{tab1}
\end{table}
Each vertex (graph) describes a social network individuals (data and metadata). Hence instead of considering each node as a simple integer, we consider the following mapping:
 \begin{eqnarray}
	V   & \rightarrow & \{0, 1\}^n \label{eqmap} \\ 
	v_i & \mapsto     & (v_i^1, v_i^2, \ldots, v_i^n) \nonumber
\end{eqnarray}
where $n$ is the number of features or attributes (data and metadata) and where $v_i^j = 1$ if feature $j$ is present for individual $v_i$. A correspondence table is generally defined between the data/metadata (complex structures) and their index with respect a given arbitrary order. For $n > 64$ we use GMP\footnote{\url{https://gmplib.org/}} to manipulate large integers.
\subsubsection{Graphs/Social Networks Analysis, Manipulation \& Visualisation}
There are several software programs for generating, managing, and manipulating graphs/social networks, varying in complexity. A fairly comprehensive list is provided in \cite{sna}. We have chosen to use Gephi\footnote{\url{https://gephi.org/}}.

Gephi~\cite{gephipaper} is an open source software for graph and network analysis. It uses a 3D render engine to display large networks in real-time and to speed up the exploration. Gephi allows to work with complex data sets and produce valuable visual results. It provides easy and broad access to network data and allows for spacializing, filtering, navigating, manipulating and clustering. 

Gephi supports different input format for data among which we choose GEXF as the most powerful one. GEXF (\textit{Graph Exchange XML Format}) is a language for describing complex networks structures, their associated data and dynamics. The following code illustrates the rich possibilities to code graphs/social networks.
\small
\begin{lstlisting}
	[...]
	<creator>Gephi.org</creator>
	<description>A Web network</description>
	</meta>
	<graph defaultedgetype="directed">
	<attributes class="node">
	<attribute id="0" title="url" type="string"/>
	<attribute id="1" title="indegree" type="float"/>
	<attribute id="2" title="frog" type="boolean">
	<default>true</default>
	</attribute>
	</attributes>
	<nodes>
	<node id="0" label="Gephi">
	<attvalues>
	<attvalue for="0" value="http://gephi.org"/>
	<attvalue for="1" value="1"/>
	</attvalues>
	</node>
	<node id="1" label="Webatlas">
	<attvalues>
	<attvalue for="0" value="http://webatlas.fr"/>
	<attvalue for="1" value="2"/>
	</attvalues>
	</node>
    [...]
	<edges>
	<edge source="0" target="1"/>
	<edge source="0" target="2"/>
	[...]
	</edges>
    [...]
	</gexf>
\end{lstlisting}
\subsection{Hypergraphs: One Level Higher} \label{hypergraph}
Hypergraphs are a generalization of graphs, hence many of the definitions of graphs are borrowed by hypergraphs. In this section we introduce basic notions about hypergraphs. Starting from their classic definition, we will adapt it for use as a graph of partial sub-graphs (communities). However, this remains a special case of hypergraphs. The purpose aims at considering more complex communities organization for multi-level parallel communication (Russian doll principle) based on this instance of hypergraphs. For a deep presentation of hypergraphs, the reader can refer to~\cite{berge1970,hgraph}. While hypergraphs are considering undirected links (edges) between nodes, we will consider directed relationships (dihypergraphs). But for sake of simplicity we recall here basic notions of hypergraphs and not those for dihypergraphs (see \cite[Chapter 6]{hgraph}). The generalization however is straightforward, at least in the context of the present work.

Let us consider a finite set of indices $I \subsetneq \mathbb{N}$.
\begin{definition}
A hypergraph $H$ denoted by $H = (V ; E = (e_i)_{i\in I}$ on a finite set $V$ is a family $(e_i)_{i\in I}$ of subsets of $V$ called hyperedges. 
The order of the hypergraph $H = (V; E)$ is the cardinality $|V| = n$ of $V$. The size of $H$ is the cardinality $|E| = m$ of $E$.
\end{definition}
Two vertices in a hypergraph are adjacent if there is a hyperedge which contains both vertices. Two hyperedges in a hypergraph are incident if their intersection is not empty.
 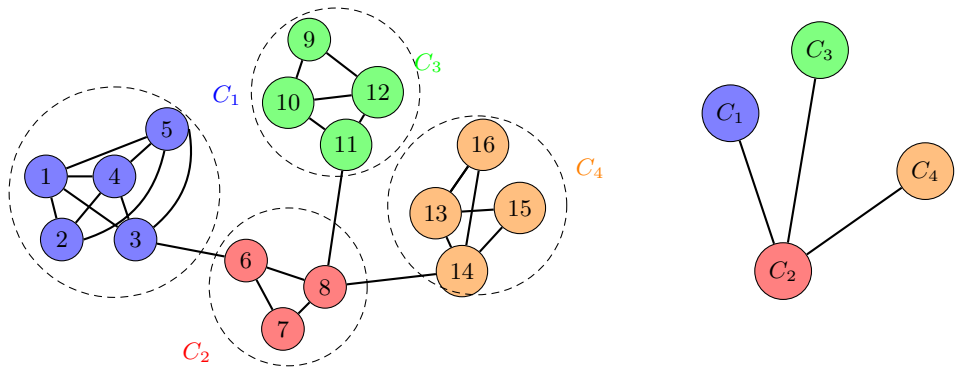
\begin{figure}[h]
 \begin{center}
 	\begin{tikzpicture}[scale=0.7]
 		\tikzstyle{ind}=[circle,draw,fill=blue!50,text=black]
 		\tikzstyle{indr}=[circle,draw,fill=red!50,text=black]
 		\tikzstyle{indg}=[circle,draw,fill=green!50,text=black]
 		\tikzstyle{indo}=[circle,draw,fill=orange!50,text=black]
 		\tikzstyle{indh}=[circle=10mm,draw,fill=blue!50,text=black]
 		\tikzstyle{indrh}=[circle=10mm,draw,fill=red!50,text=black]
 		\tikzstyle{indrg}=[circle=10mm,draw,fill=green!50,text=black]
 		\tikzstyle{indro}=[circle=10mm,draw,fill=orange!50,text=black]
 		\tikzstyle{lien01}=[-,thick]
 		\tikzstyle{lien11}=[-,thick]
 		\node[ind] (1) at (-6,3.3) {$1$};
 		\node[ind] (2) at (-5.7,2.1) {$2$};
 		\node[ind] (3) at (-4.3,2.1) {$3$};
 		\node[ind] (4) at (-4.7,3.3) {$4$};
 		\node[ind] (5) at (-3.7,4.2) {$5$};
 		\draw[lien01] (1)--(2);
 		\draw[lien01] (1)--(4);
 		\draw[lien11] (2)--(4);
 		\draw[lien11] (1)--(3);
 		\draw[lien01] (1)--(5);
 		\draw[lien11] (4)--(5);
 		\draw[lien01] (5)to[bend left](2.east);
 		\draw[lien01] (3)to[bend right](5.east);
 		\draw[lien11] (4)--(3);
 		\draw[densely dashed] (-4.7,3.0) circle[radius=2.0cm];
 		\node[above right] at (-3.0,4.5) {\color{blue}{$C_1$}};
 		\node[indr] (6) at (-2.2,1.7) {$6$};
 		\node[indr] (7) at (-1.5,0.4) {$7$};
 		\node[indr] (8) at (-0.7,1.2) {$8$};
 		\draw[lien01] (6)--(8);
 		\draw[lien11] (6)--(7);
 		\draw[lien11] (7)--(8);
 		\draw[lien01] (6)--(3);
 		\draw[densely dashed] (-1.4,1.2) circle[radius=1.5cm];
 		\node[below left] at (-2.7,0.3) {\color{red}{$C_2$}};
 		\node[indg] (9) at (-1.0,5.9) {$9$};
 		\node[indg] (10) at (-1.4,4.7) {$10$};
 		\node[indg] (11) at (-0.3,3.9) {$11$};
 		\node[indg] (12) at (0.3,4.9) {$12$};
 		\draw[lien01] (9)--(10);
 		\draw[lien01] (9)--(12);
 		\draw[lien11] (10)--(12);
 		\draw[lien11] (11)--(12);
 		\draw[lien01] (11)--(10);
 		\draw[lien01] (8)--(11);
 		\draw[densely dashed] (-0.5,4.9) circle[radius=1.6cm];
 		\node[below left] at (1.7,5.8) {\color{green}{$C_3$}};
 		\node[indo] (13) at (1.4,2.6) {$13$};
 		\node[indo] (14) at (1.9,1.5) {$14$};
 		\node[indo] (15) at (3.0,2.7) {$15$};
 		\node[indo] (16) at (2.3,3.9) {$16$};
 		\draw[densely dashed] (2.19,2.75) circle[radius=1.7cm];
 		\node[below right] at (3.9,3.8) {\color{orange}{$C_4$}};
 		\draw[lien01] (16)--(14);
 		\draw[lien01] (13)--(15);
 		\draw[lien11] (14)--(13);
 		\draw[lien11] (13)--(16);
 		\draw[lien01] (13)--(16);
 		\draw[lien01] (14)--(15);
 		\draw[lien01] (8)--(14);
 		
 		\node[indh] (C1) at (7.0,4.5) {$C_1$};
 		\node[indrh] (C2) at (8.0,1.5) {$C_2$};
 		\node[indrg] (C3) at (8.7,5.7) {$C_3$};
 		\node[indro] (C4) at (10.7,3.4) {$C_4$};
 		\draw[lien01] (C2)--(C1);
 		\draw[lien01] (C2)--(C3);
 		\draw[lien01] (C2)--(C4);
 	\end{tikzpicture}
 \end{center}
 \caption{Graphs $G$ with 4 communities (left) - Hypergrah view of $G$} \label{hyper4} 
 \end{figure}
In the context of social network communities, we can consider one level higher where communities, including adjacent elements of other communities are hyperedges. Somehow and conveniently, this particular vision enables to model communities (hyperedges) as meta-nodes and edges (arcs) between meta-nodes describes connectivity between communities. Figure~\ref{hyper4} illustrates this with $V = \{1, 2, 3, 4, 5, \ldots 16\}$ and $ E = \{e_1, e_2, e_3, e_4\}$ with $e_1 = \{1, 2, 3, 4, 5, 6\}, e_2 = \{3, 6, 7, 8, 11\}, e_3 = \{8, 9, 10, 11, 12\}$ and $e_4 = \{8, 13, 14, 15, 16\}$. 

\section{Presentation of the Community Covert Communication Technique (CCCT)} \label{s3}
\subsection{Working Assumptions}
From the field data and cases presented in Section~\ref{s1}, we fix the following working assumptions for the Community Covert Communication technique:
\begin{itemize}
	\item Actors: one sock puppet master $P$ who intends to communicate with $\alpha$ users (without loss of generality we will consider $\alpha = 2$). We call these users ``CCC users''.
	\item Average sock puppet community size $N = 50,000$.  
	\item $P$ manages a finite number $c$ of communities either in a unique social network or spilled up on different social networks (Twitter/X, LinkedIn, Facebook, Telegram, TikTok, Amazon, Airbnb \ldots). Each community is made up of sock puppet accounts and of the $\alpha$ user's profiles.
	\item Each user has access to a unique API with two secret keys. A unique secret key $K_0$ and a user-dependent secret key $K'_i$ with $1 \leq i \leq \alpha$.
	\item An API is provided by $P$ to the $\alpha$ users. It can be updated frequently. The API is supposed to be secret, but its secrecy does not determine the security of the CCC protocol. The API enables to identify specific communities, to extract data from communities' profiles and to translate into a ciphertext. 
\end{itemize}
We adopt the following notation for the rest of the paper. Let $H$ be the SHA2 hash function. Then we note $H^k(X, K) = H(H^{k - 1}(X, K))$ with $H^1(X, K) = H(X, K)$.
\subsection{The General Principle of CCCT}
In any social network, atomic entities are individuals with a profile containing data and metadata. According to the Mapping~(\ref{eqmap}), each individual $v_i$ is described by the $n$-uple $(v_i^1, v_i^2, \ldots, v_i^n)$.
\subsubsection{Preliminaries}
First, $P$ generates and inserts new individuals (\textit{personae} or \textit{sock puppets}) within the target social network and then dynamically animates/manages them. He can create, delete, and modify quantitatively and qualitatively as many links as he wishes between real individuals in the network and sock puppet accounts. The goal is twofold: on the one hand, to ``drown'' them within the existing social network and, on the other hand, to still be able to easily and quickly identify these individuals and only these. 

It should be noted that while $P$ organizes these $N$ individuals into a community, including the $\alpha$ users, no other person should be able to isolate them as a constituted community, except the $\alpha$ users, provided that they have the correct secret key $K_0$. To achieve this, each individual (sock puppet accounts and CCC users) must be ``semantically and socio-semantically neutral'', unlike the profiles created and used, in essence, for the purpose of influence.

The API provided to CCC users enables:
\begin{itemize}
\item to identify the target community members precisely,
\item to validate each sock puppet account,
\item to select a subset of sock puppet accounts that is used to encode the current ciphertext $C$. 
\end{itemize}
For the validation part, we use (among many other possibilities), Bloom filters~\cite{bloom1970} in an optimized way~\cite{filiol-bloom}. The array size for the filter is $T = 2^{20}$ and we use $k = 64$ hash functions of type MurmurHash3~\cite{appleby}. The secret key $K_0$ is used as seed in these hash functions (in fact we use $H^3(\mathcal{N}, K_0)$ where $\mathcal{N}$ is a (public or not) nonce\footnote{In our case we used a random, unique 512-bit nonce coding the communication instance.}). Data are made of the $n$-tuple $(v_i^1, v_i^2, \ldots, v_i^n)$ for each user $v_i$.   
\subsection{Selection of a Community Support Subset}
From the target community of size $N$, CCC users have to select a subset of sock puppet accounts, randomly and thanks to the API. 
Let us consider a community $\mathcal{C} = \{v_i\}_{1 \leq i \leq N}$ and a ciphertext to encode of size $|C|$. We note $C = c_1, c_2, \ldots, c_{|C|}$. 

For the encoding step we need to extract a subset $\mathcal{S}$ of size $s < N$ from community $\mathcal{C}$. The API is then using a stream cipher $E$ initialized with $H(\mathcal{N}, K_0)$ to select this subset as follows:
\begin{enumerate}
\item The API generates a pseudo-random sequence $(\sigma_i)_{1 \leq i \leq 2.s} = E(H(\mathcal{N}, K_0))$
\item Then a decimation process is applied to select $\mathcal{S}$ from $\mathcal{C}$ such that 
\[
\left\{
\begin{array}{ll}
	\mbox{if } \sigma_i = 1 & \mbox{$v_i$ from $\mathcal{C}$ is kept in $\mathcal{S}$} \\
	\mbox{otherwise }       & \mbox{$v_i$ from $\mathcal{C}$ is discarded} 
\end{array}
\right.
\]
We then have $\mathcal{S} = \{v_{i_1}, v_{i_2}, v_{i_3}, \ldots, v_{i_s}\}$.
\end{enumerate}
\subsubsection{Ciphertext Encoding}
The core principle is that each bit of ciphertext is encoded by links between profiles of sock puppet accounts. 

For any two users $(v_i^1, v_i^2, \ldots, v_i^n)$ and $(v_j^1, v_j^2, \ldots, v_j^n)$ $P$ can set a link between $(v_i)$ and $(v_j)$ (they are somehow connected in a way or another, using profile attributes) or make $(v_i)$ and $(v_j)$ not connected with respect to the same attributes. The API provided to CCC users is able to extract connections whether they exist or not. 

Without loss of generality, we will not take loops into account, whether the graph is directed or not. In a social network, this corresponds to a self-connected or self-referenced profile. 
Starting from the subset community $\mathcal{S} = \{v_{i_1}, v_{i_2}, v_{i_3}, \ldots, v_{i_s}\}$, the API selects an arbitrary order for possible links between nodes, randomly. The random generator is initialized by the value $H^2(\mathcal{N}, K_0)$. Let us suppose that we have $i_1 < i_2 < i_3 < \ldots < i_s$, the trivial order is given by
\begin{itemize}
\item $\{i_1, i_2\}, \{i_1, i_3\}, \ldots \{i_1, i_s\}, \{i_2, i_3\}, \{i_2, i_4\}, \ldots, \{i_2, i_s\}, \ldots \{i_{s - 1}, i_s\}$ for undirected graph. There are $\frac{s.(s - 1)}{2}$ such links.
\item $(i_1, i_2), \ldots (i_1, i_3), \ldots, (i_1, i_s), (i_2, i_1), (i_2, i_3) \ldots, (i_{s - 1}, i_1), \ldots (i_{s - 1}, i_s)$ for a directed graph. There are $s.(s - 1)$ such directed links.
\end{itemize} 
The arbitrary random link order is performed by using a $K_0$-dependent random, bijective permutation $\tau$ of the indices. For instance, starting from the trivial link order for undirected graph we finally get the following link order 
\begin{equation*}
\begin{split}
& e_1 = \{\tau(i_1), \tau(i_2)\}, e_2 = \{\tau(i_1), \tau(i_3)\}, \ldots e_{s - 1} = \{\tau(i_1), \tau(i_s)\}, e_s =  \{\tau(i_2), \tau(i_3)\}, \\
&  e_{S + 1} = \{\tau(i_2), \tau(i_4)\}, \ldots,e_{2. s - 3} = \{\tau(i_2), \tau(i_s)\}, \ldots e_i = \{\tau(i_{s - 1}), \tau(i_s)\} \\
\end{split}
\end{equation*}

Finally $P$ sets up links between sock puppet accounts, with respect to given attributes, to encode the ciphertext $C = c_1, c_2, \ldots, c_{|C|}$ as follows:
\[
\left\{
\begin{array}{ll}
	\mbox{if $c_i = 1$}  & \mbox{link $e_i$ is present in $\mathcal{S}$} \\
	\mbox{otherwise }    & \mbox{link $e_i$ is absent in $\mathcal{S}$} 
\end{array}
\right.
\]
As for the CCC users, the API, with the correct secret key $K_0$ enables to scrap data and conversely translates links within sub-community $\mathcal{S}$ as ciphertext $C$.
\subsubsection{Encoding Rate}
Depending on whether the graph (links in the corresponding social network) is directed or undirected, whether we accept loops or not, the encoding rate is summarized in Table~\ref{tab-rate} when considering a sub-community $\mathcal{S}$ of size $s$.

\begin{figure}[h]
\begin{center} 
	\includegraphics[width=0.9\textwidth]{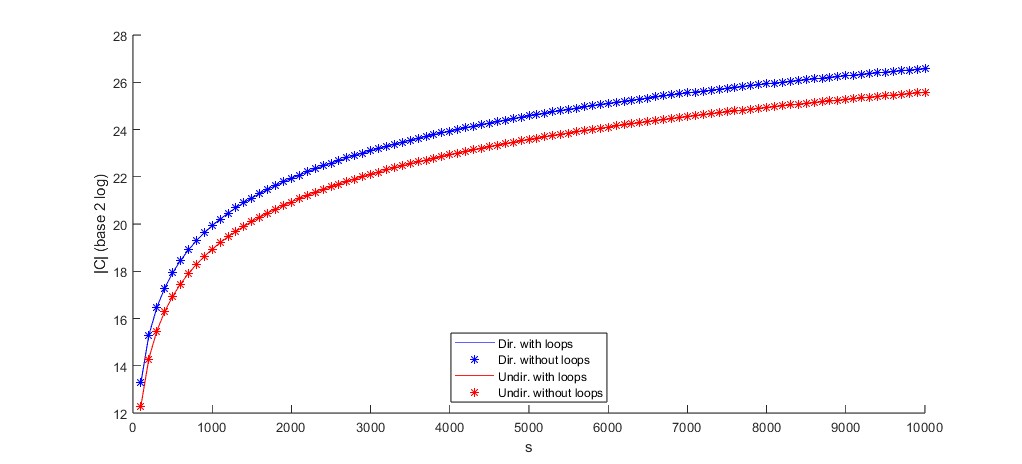}
\end{center}
\caption{Number of ciphertext bits (base 2 log) depending on sub-community size $s$} \label{s2mvar}
\end{figure}

Figure~\ref{s2mvar} shows the size of ciphertext we can encode depending on the sub-community size $s$ for the four cases in Table~\ref*{tab-rate}.\\
\begin{table}[h!]
	\centering
	\begin{tabular}{|l|c|c|} \hline
		Graph/social network type & Encoding capability (in bits) & $|C|$ for $s = 5,000$   \\ \hline \hline
		Directed with loops       &   $s^2$                       &       $\approx$ 24~Mb   \\
		Directed without loops    &   $s.(s - 1)$                 &       $\approx$ 24~Mb   \\ 
		Undirected with loops     &   $\frac{s.(s - 1)}{2} + n$   &       $\approx$ 12~Mb   \\
		Undirected without loops  &   $\frac{s.(s - 1)}{2}$       &       $\approx$ 12~Mb   \\ \hline
	\end{tabular}
	\caption{Encoding rate with respect to graph/social network nature} \label{tab-rate}
\end{table}
\raggedbottom
\subsection{An Illustrative Example}
Let us consider the following example to illustrate things very simply. We want to encode the following 36-bit ciphertext 
\[C = 110100100011100100101011010011101011\]
For these two examples, we suppose that we directly work on subset $\mathcal{S}$.
\begin{figure}
	\centering
	\scalebox{.85}{
		\begin{tabular}{cc}
			\begin{tikzpicture}[scale=1.0]
				\tikzstyle{ind}=[circle,draw,fill=white!50,text=black]
				\tikzstyle{lien00}=[-,thick]
				\node[ind] (1) at (0:3cm) {$1$};
				\node[ind] (2) at (40:3cm) {$2$};
				\node[ind] (3) at (80:3cm) {$3$};
				\node[ind] (4) at (120:3cm) {$4$};
				\node[ind] (5) at (160:3cm) {$5$};
				\node[ind] (6) at (200:3cm) {$6$};
				\node[ind] (7) at (240:3cm) {$7$};
				\node[ind] (8) at (280:3cm) {$8$};
				\node[ind] (9) at (320:3cm) {$9$};
				\draw[lien00] (1)--(2);
				\draw[lien00] (1)--(3);
				\draw[lien00] (1)--(5);
				\draw[lien00] (1)--(8);
				\draw[lien00] (2)--(5);
				\draw[lien00] (2)--(6);
				\draw[lien00] (2)--(7);
				\draw[lien00] (3)--(4);
				\draw[lien00] (3)--(7);
				\draw[lien00] (3)--(9);
				\draw[lien00] (4)--(6);
				\draw[lien00] (4)--(7);
				\draw[lien00] (4)--(9);
				\draw[lien00] (5)--(8);
				\draw[lien00] (5)--(9);
				\draw[lien00] (6)--(7);
				\draw[lien00] (6)--(9);
				\draw[lien00] (7)--(9);
				\draw[lien00] (8)--(9);
				
			\end{tikzpicture} &
			\begin{tikzpicture}[scale=1.0]
				\tikzstyle{ind}=[circle,draw,fill=white!50,text=black]
				\tikzstyle{lien01}=[->,thick]
				\tikzstyle{lien11}=[<->,thick]
				\node[ind] (1) at (0:3cm) {$1$};
				\node[ind] (2) at (42:3cm) {$2$};
				\node[ind] (3) at (94:3cm) {$3$};
				\node[ind] (4) at (146:3cm) {$4$};
				\node[ind] (5) at (198:3cm) {$5$};
				\node[ind] (6) at (250:3cm) {$6$};
				\node[ind] (7) at (302:3cm) {$7$};
				
				\draw[lien11] (1)--(2);
				\draw[lien11] (1)--(3);
				\draw[lien01] (1)--(5);
				\draw[lien01] (2)--(6);
				\draw[lien01] (2)--(7);
				\draw[lien01] (3)--(5);
				\draw[lien01] (4)--(1);
				\draw[lien01] (4)--(3);
				\draw[lien01] (4)--(6);
				\draw[lien01] (4)--(7);
				\draw[lien01] (5)--(2);
				\draw[lien11] (5)--(6);
				\draw[lien01] (5)--(7);
				\draw[lien01] (6)--(1);
				\draw[lien01] (6)--(3);
				\draw[lien01] (6)--(7);
				
			\end{tikzpicture}
	\end{tabular}}
	\caption{Coding communities for 36-bit ciphertext $C$ with undirected graph (left) and directed graph (right)} \label{coding}
\end{figure}
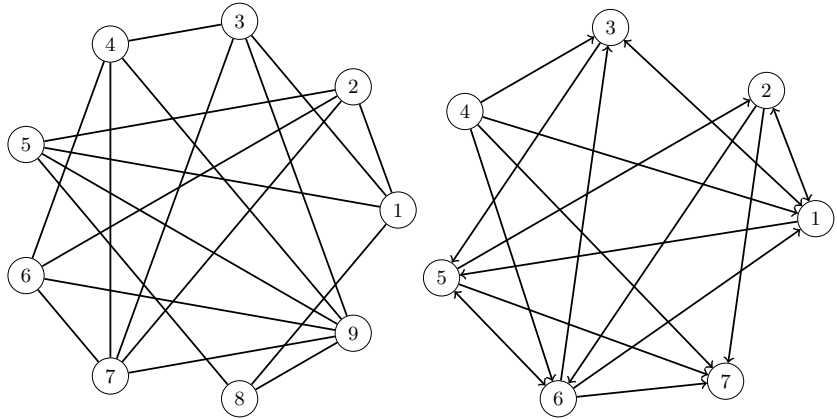
If we consider undirected (social/graph) links, we need a coding community of $N = 9$ nodes, while for directed (social/graph) links we just need $N = 7$ nodes (for sake of simplicity we do not consider loop; considering loops enable to need only $N = 6$ nodes). 
\subsubsection{Undirected social links} By means of $K_0$ we define an arbitrary reference order for undirected links between $N = 9$ nodes. For sake of simplicity we consider the trivial order. Then we associate a ciphertext bit $c_i$ to each undirected link, with respect to this arbitrary order. Then we have
\[
\begin{array}{l||cccccccccccc}
\textbf{Links} & \{1,2\} & \{1,3\} & \{1,4\} & \{1,5\} & \{1,6\} & \{1,7\} & \{1,8\} & \{1,9\} & \{2,3\} & \{2,4\} & \{2,5\} & \{2,6\} \\
\mathbf{c_i}   &    1    &    1    &    0    &    1    &    0    &    0    &    1    &    0    &    0    &    0    &    1    &    1    \\ \hline
               &         &         &         &         &         &         &         &         &         &         &         &         \\ 
\textbf{Links} & \{2,7\} & \{2,8\} & \{2,9\} & \{3,4\} & \{3,5\} & \{3,6\} & \{3,7\} & \{3,8\} & \{3,9\} & \{4,5\} & \{4,6\} & \{4,7\} \\
\mathbf{c_i}   &    1    &    0    &    0    &    1    &    0    &    0    &    1    &    0    &    1    &    0    &    1    &    1    \\ \hline
               &         &         &         &         &         &         &         &         &         &         &         &         \\
\textbf{Links} & \{4,8\} & \{4,9\} & \{5,6\} & \{5,7\} & \{5,8\} & \{5,9\} & \{6,7\} & \{6,8\} & \{6,9\} & \{7,8\} & \{7,9\} & \{8,9\} \\
\mathbf{c_i}   &    0    &    1    &    0    &    0    &    1    &    1    &    1    &    0    &    1    &    0    &    1    &    1    \\ \hline
\end{array}
\] 
The resulting coding community is depicted in Figure~\ref{coding} (left).
\begin{figure}[h!]
	\centering
		\begin{tabular}{lll}
			\includegraphics[width=3.75cm]{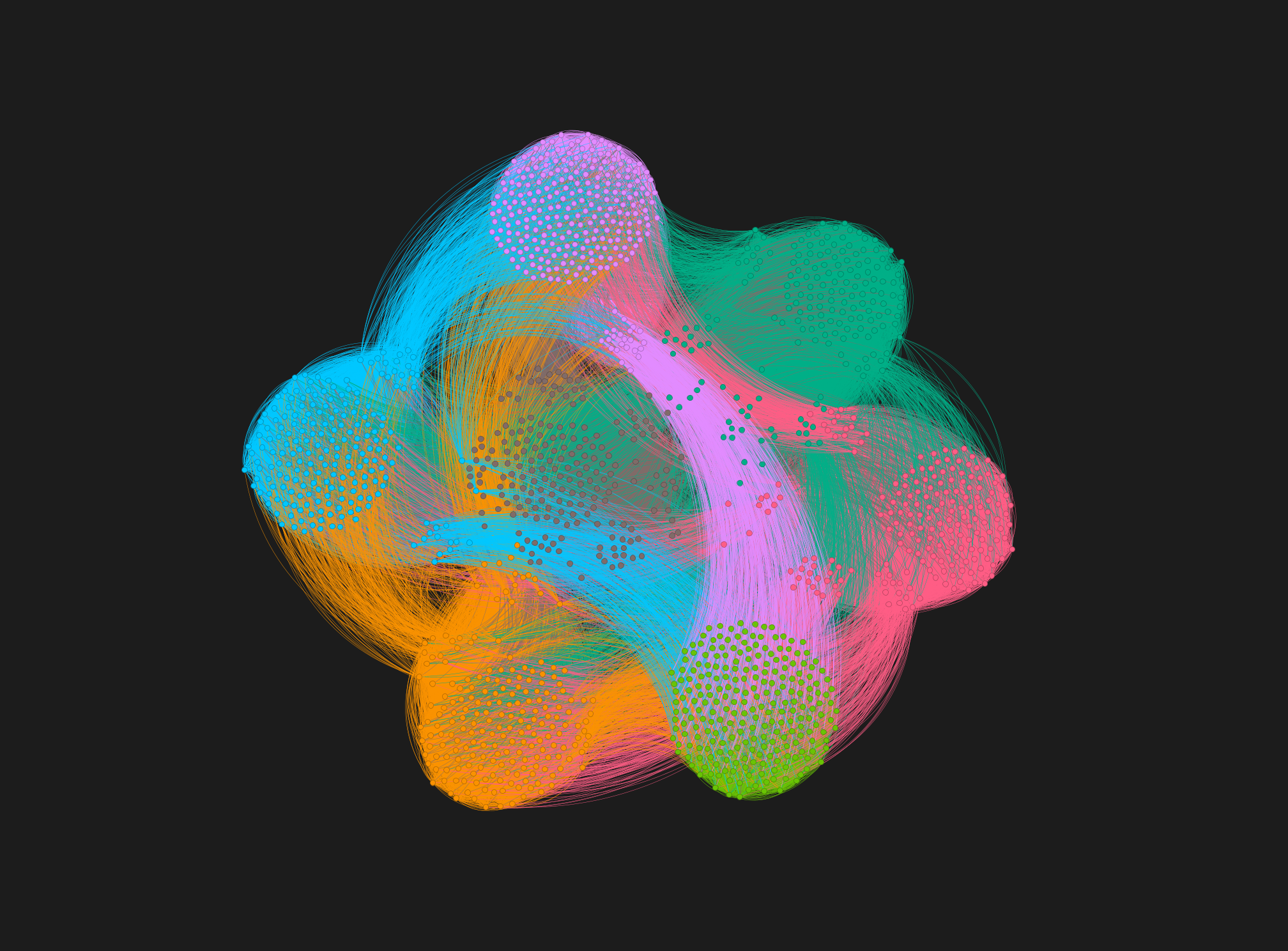} & \includegraphics[width=3.750cm]{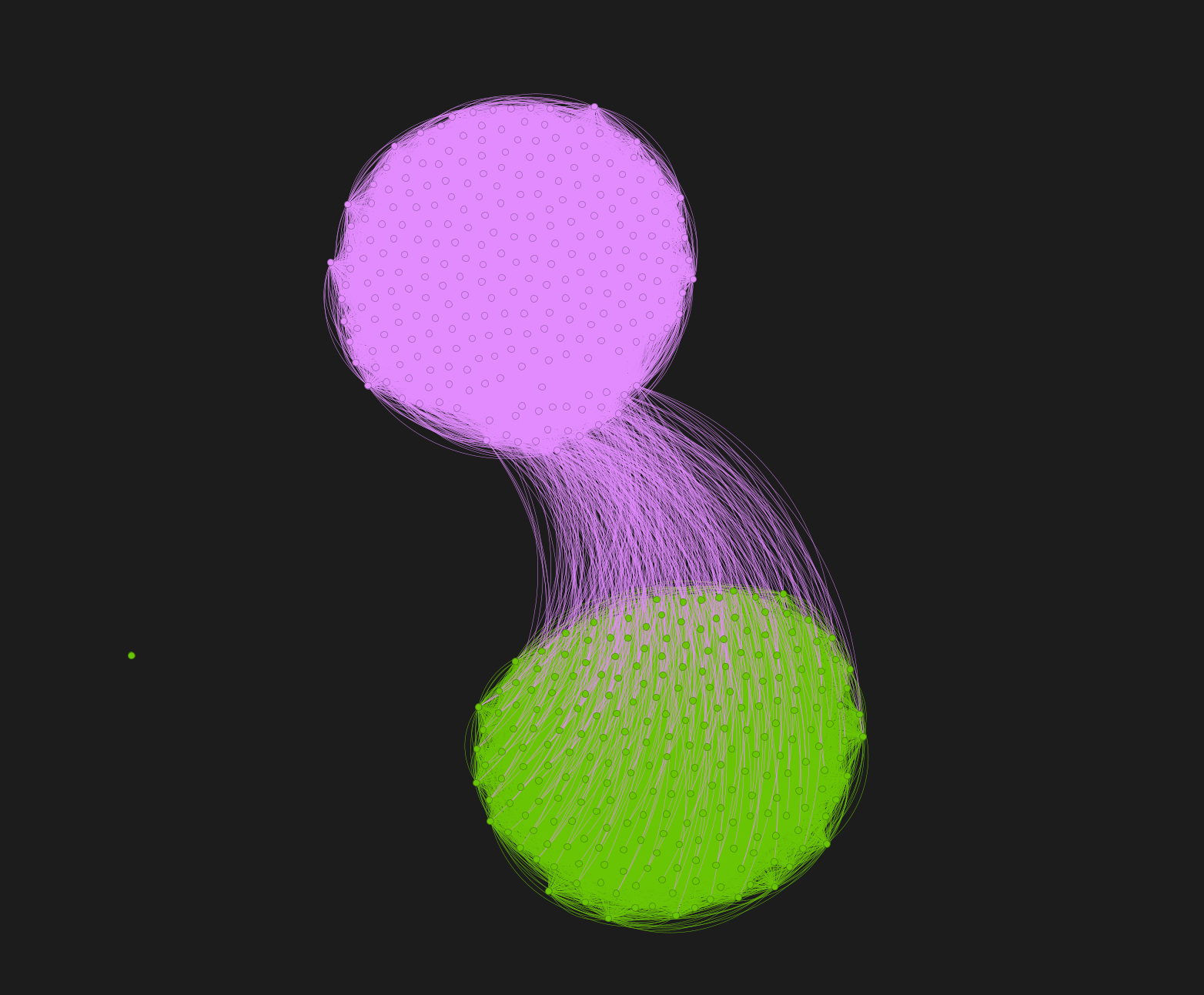} & \includegraphics[width=3.750cm]{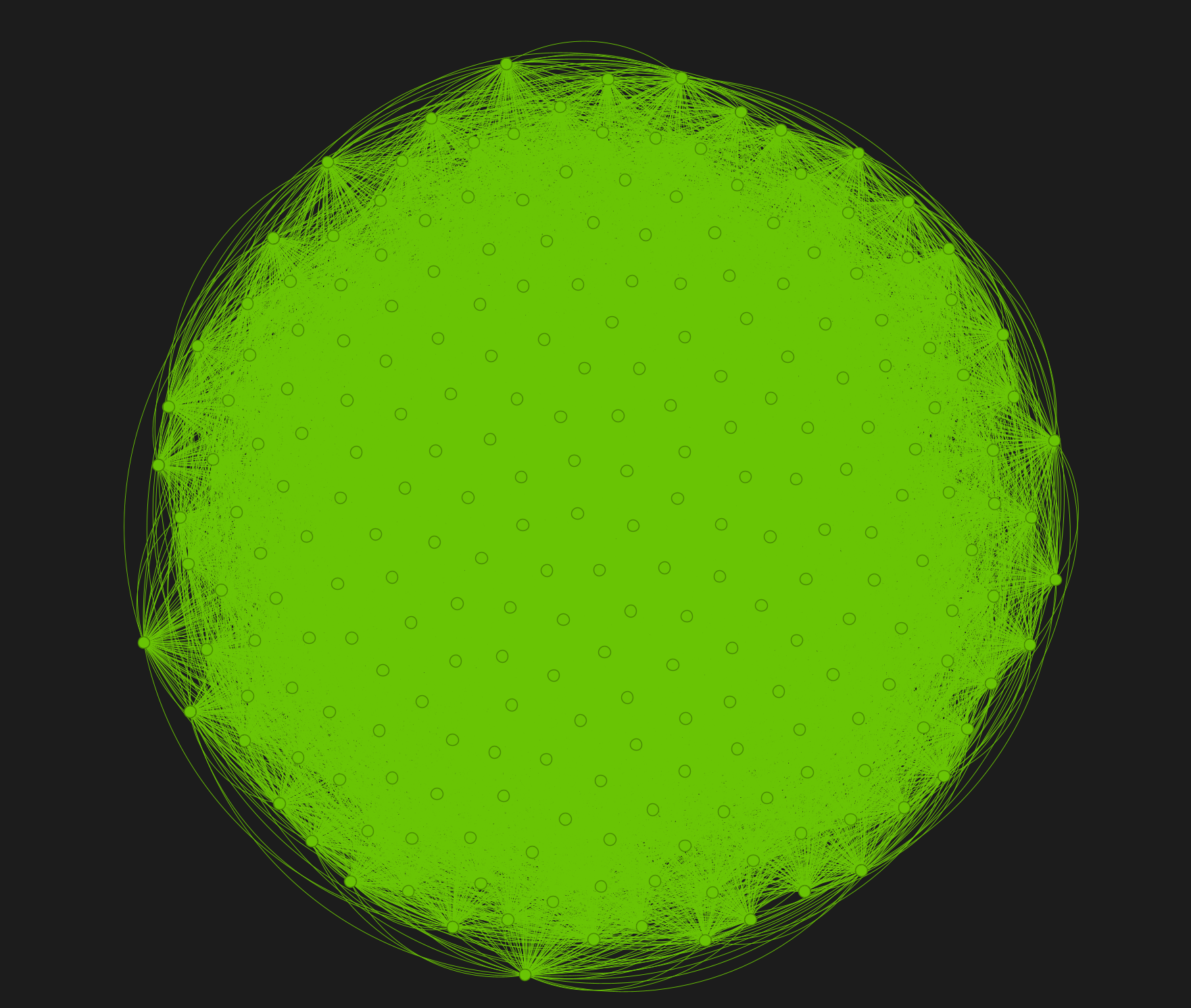}
		\end{tabular}
	\caption{Gephi graph (Fruchterman Reingold layout)for a 4 Kb ciphertext (left, all nodes; center, community nodes; right, sub-community encoding nodes only)} \label{real}
\end{figure}
\raggedbottom

We have encoded a ciphertext of 4Kb (the one used for 44CON 2024 demos for non trivial deniable cryptography). Figure~\ref{real} shows the corresponding Gephi graphs obtained (1,489 nodes among which we have $S = 237$ encoding nodes selected in a 357-node sock puppet community and a total of 94,547 edges).
\subsubsection{Directed social links} In the same way, by means of $K_0$ we define an arbitrary reference order for directed links between $N = 7$ nodes. For sake of simplicity we consider the trivial order. Then we similarly associate a ciphertext bit $c_i$ to each directed link, with respect to this arbitrary order. Then we have
\[
\begin{array}{l||cccccccccccc}
\textbf{Links} & (1, 2) & (1, 3) & (1, 4) & (1, 5) & (1, 6) & (1, 7) & (2, 1) & (2, 3) & (2, 4) & (2, 5) & (2, 6) & (2, 7) \\	
\mathbf{c_i}   &    1   &    1   &    0   &    1   &    0   &    0   &    1   &    0   &    0   &    0   &    1   &    1   \\ \hline
               &        &        &        &        &        &        &        &        &        &        &        &        \\ 
\textbf{Links} & (3, 1) & (3, 2) & (3, 4) & (3, 5) & (3, 6) & (3, 7) & (4, 1) & (4, 2) & (4, 3) & (4, 5) & (4, 6) & (4, 7) \\
\mathbf{c_i}   &    1   &    0   &    0   &    1   &    0   &    0   &    1   &    0   &    1   &    0   &    1   &    1   \\ \hline
               &        &        &        &        &        &        &        &        &        &        &        &        \\
\textbf{Links} & (5, 1) & (5, 2) & (5, 3) & (5, 4) & (5, 6) & (5, 7) & (6, 1) & (6, 2) & (6, 3) & (6, 4) & (6, 5) & (6, 7) \\
\mathbf{c_i}   &    0   &    1   &    0   &    0   &    1   &    1   &    1   &    0   &    1   &    0   &    1   &    1   \\ \hline
\end{array}	
\]
The resulting coding community is depicted in Figure~\ref{coding} (right).	
\subsection{One Level Up - Using Hypergraphs}
Considering hypergraphs (see Section~\ref{hypergraph}), $P$ can encode encrypted texts with links between communities, parallel to the links within each community. To illustrate this, referring to Figure~\ref{hyper4}, $P$ can therefore send 5 ciphertexts at the same time (with respect to the trivial order of the links) according to the following table:

\begin{table}[h]
\scalebox{0.80}{
\begin{tabular}{c||cccccccccc}
	\textbf{Links $C_1$} & (1, 2) & (1, 3) & (1, 4) & (1, 5) & (2, 3) & (2, 4) & (2, 5) & (3, 4) & (3, 5) & (4, 5) \\	
	\textbf{Ciphertext 1}&    1   &    1   &    1   &    1   &    0   &    1   &    1   &    1   &    1   &    1   \\ \hline
	                     &        &        &        &        &        &        &        &        &        &        \\ 
	\textbf{Links $C_2$} & (6, 7) & (6, 8) & (8, 9) &        &        &        &        &        &        &        \\
	\textbf{Ciphertext 2}&    1   &    1   &    1   &        &        &        &        &        &        &        \\ \hline
	                     &        &        &        &        &        &        &        &        &        &        \\
	\textbf{Links $C_3$} & (9, 10) & (9, 11) & (9, 12) & (10, 11) & (10, 12) & (11, 12) &        &        & &       \\
	\textbf{Ciphertext 3}&    1   &    0   &    1   &    1   &    1   &    1   &    1   &        &        &        \\ \hline
	                     &        &        &        &        &        &        &        &        &        &        \\
	\textbf{Links $C_4$} & (13, 14) & (13, 15) & (13, 16) & (14, 15) & (14, 16) & (15, 16) &     &        &   &     \\
	\textbf{Ciphertext 4}&    1   &    1   &    1   &    1   &    1   &    1   &    0   &        &        &        \\ \hline
	                     &        &        &        &        &        &        &        &        &        &        \\
\textbf{Hypergraph links}& ($C_1, C_2)$ & $(C_1, C_3)$ & $(C_1, C_4)$ & $(C_2, C_3)$ & $(C_2, C_4)$ & $(C_4, C_4)$ &  & & &    \\ 
    \textbf{Ciphertext 5}&    1       &      0     &    0       &      1     &      1     &      0   &    &        &        &        \\ \hline
\end{tabular}	}
\end{table}
\raggedbottom
\subsection{Non Trivial Deniable Cryptography}
\begin{table}[ht!]
	\centering
	\begin{tabular}{cc}
		\includegraphics[width=6.0cm]{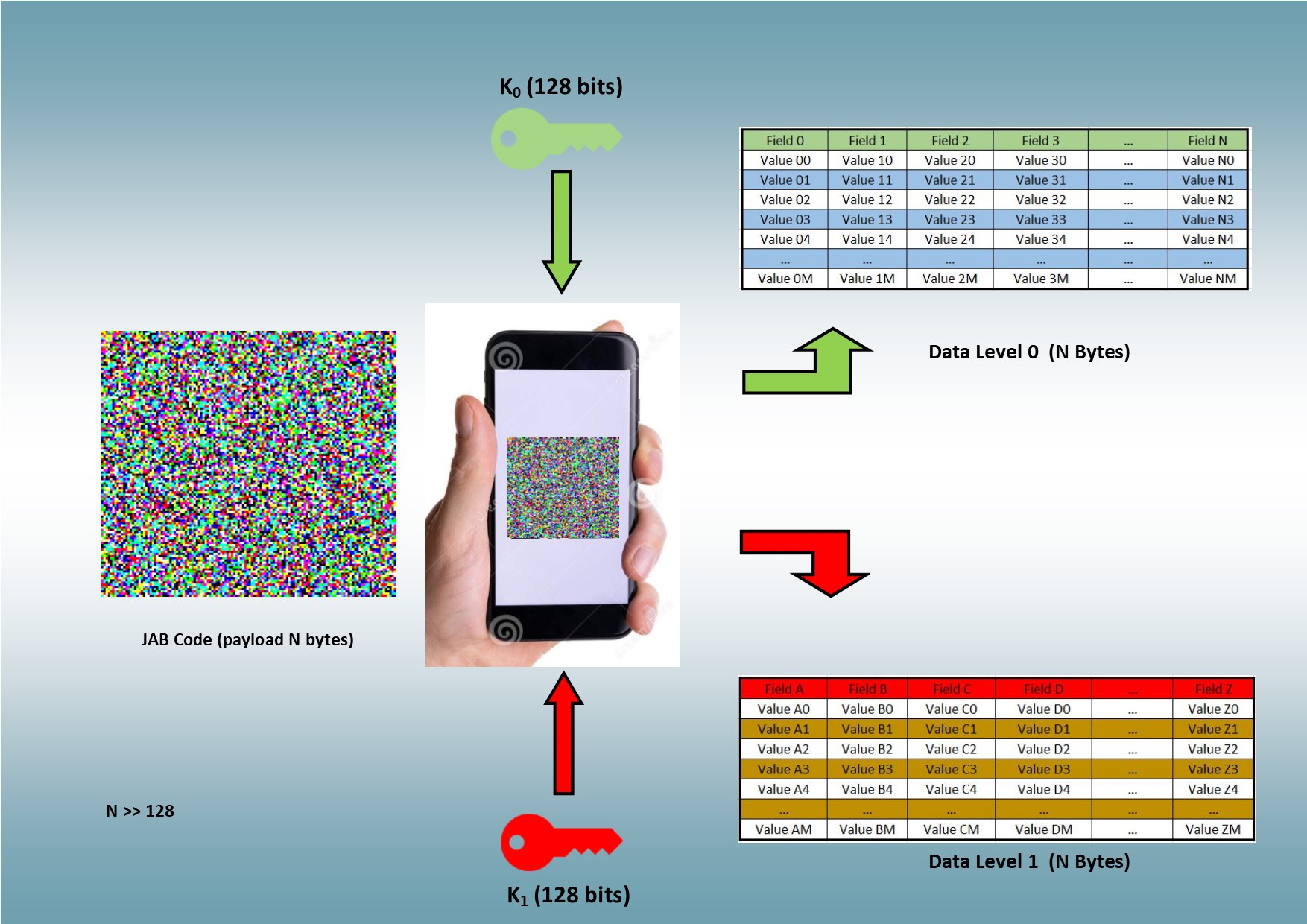} & \includegraphics[width=6.0cm]{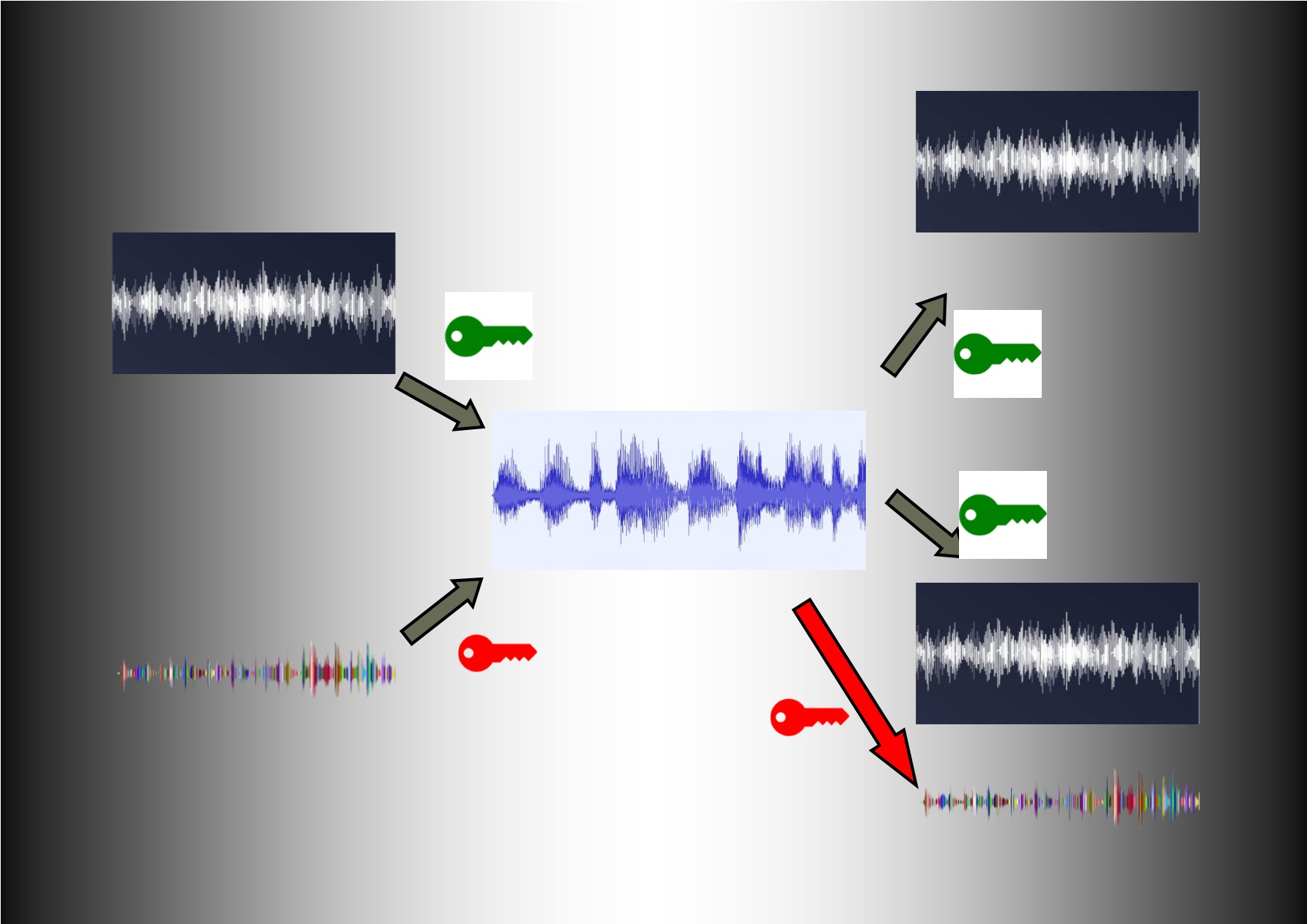}
	\end{tabular}
	\caption{Applications of non trivial deniable cryptography (left, DC-based Static Multilevel Communication. Right, DC-based Dynamic Multilevel Communication). By courtesy of Hope4Sec, \url{https://hope4sec.eu/pages/multi-level-data-protection.html}} \label{appdc}
\end{table}
This section just recalls the basics and principle of the non trivial deniable cryptography we have used. Technical details and demos have been presented at~\cite{filiol44con}. This technology has been been developed in 2021 in collaboration with Hope4Sec, Tallinn.

Effective (non trivial) deniable cryptography until very recently was an open problem. Only trivial case is known under name \textit{One Time Pad} (OTP). In this case, ``keys'' are as long as the two (or more) plaintexts and hence OTP not a valid solution (OTPs must be hardcoded).
\begin{figure}[h!]
	\centering
		\includegraphics[width=09.0cm]{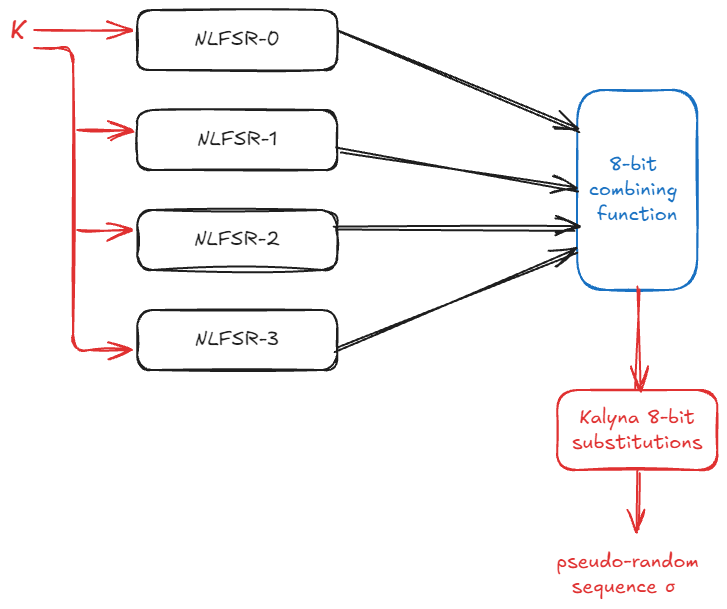} 
	\caption{General design of non trivial deniable cryptography algorithm} \label{dcgen}
\end{figure}
\raggedbottom

Let $C$ be a ciphertext of length $|C| = N$, a \textbf{unique} algorithm $E$ and any two different arbitrary plaintexts $P_1$ and $P_2$. We have developed a C framework to build encryption algorithms for effective deniable cryptography with short keys (128 - 256 bits) on the fly.
\begin{itemize}
\item $E$ is deterministic and resistant to known cryptanalysis techniques. We use Non Linear Feedback Shift Registers (NLFSRs) combined by a highly non linear Boolean Function (see Figure~\ref{dcgen}). 
\item Keys are $k$-bit long, with $k$ far smaller than $N$ (so one-time pad is not considered). Generally we use 128- and 256-bit secret keys.
\item We have $C = E(K_1, P_1) = E(K_2, P_2)$ (extendable to a finite number of plaintexts $P_i$) \textbf{AND} \color{red}{$|P_i| = |C|$} \color{black}{for all $i$}.
\end{itemize}
Different applications have been developed and currently are by Hope4Sec (Figure~\ref{appdc}). For instance, non-trivial DC can be used, for example, to protect databases by encrypting them and providing two levels of data criticality depending on the secret key used (Figure~\ref{appdc}, left). Another application allows two encrypted communications to be transmitted on the same stream, with either secret key giving access to two different levels of communication (Figure~\ref{appdc}, right).
\subsection{Applications and Use-cases}
In addition to encrypted content covert communication, we already have identified additional applications and are working on new ones, among which are
\begin{itemize}
\item coding a number of warrant canaries~\cite{warrantcanary} or activation messages,
\item distributing cryptographic keys,
\item implementing dead man switch mechanisms~\cite{dms},
\item countdown mechanisms,
\item data storage,
\item \ldots
\end{itemize}

\section{Security Analysis Outline of CCCT} \label{s4}
This section presents an outline of the security analysis of the CCC protocol. As with any new system, security can only be achieved over time and with the input of external researchers. 

The main advantage is that the CCC technique no longer relies on a single channel exchange. The protocol is completely decentralized. There is no longer a single source of information, but rather a multitude of sources that must be processed in parallel (every sock puppet account).
\subsection{Detection Part}
The goal is to identify one or more communities of type $\mathcal{C}$, then the sub-community $\mathcal{S}$. Even if the API provided by $P$ to CCC users is secret, its secrecy must not determine the security of the protocol. We can therefore reasonably assume that this API is analysed by an adversary. However since the API is frequently updated for any new communication instance, an adversary would have to have access to it and again perform most of the analysis again.
\begin{itemize}
	\item Identifying $\mathcal{S}$ requires the knowledge of $\mathcal{C}$ and the 256-bit secret key $K_0$. The encryption algorithm $E$ can be changed each time the API is updated. Note that the cryptanalysis of $E$ does not reveal $K_0$ but rather $H(\mathcal{N}, K_0)$.
	\item Identifying $\mathcal{C}$ also requires the knowledge of the secret key $K_0$. The Bloom filter can also be changed each time the API is updated. This filter allows $\mathcal{C}$ to be selected without necessarily knowing all the necessary data (clueless protocol). Any cryptanalysis of the Bloom filter only allows $H^3(\mathcal{N}, K_0)$ to be recovered.
	\item If the adversary attempts an approach without knowing the API, based solely on the assumption of the existence of this protocol and communities of type $\mathcal{C}$, this is a brute force approach. For this approach to be considered unrealistic and unachievable, it is essential that the content of each sock puppet account profile be ``neutral'' or insignificant. Moreover and above all, the dynamics of creating these accounts must remain under the radar and undetectable (for example, not creating thousands of accounts at once). This initial phase is the most critical, but once it is complete, it is easy to continuously modify/update existing accounts.
\end{itemize}
Community detection is still an open problem. The very definition of a community is still unclear~\cite{comm_detect}. First, it is important to bear in mind that the concept of community is, by definition, subjective. It is defined in relation to a subset $I\subsetneq \mathbb{N}^n$ of attributes (from Equation~(\ref{eqmap}). For example, the community of fans of a given television series is not the same overall as the groupies of a rock star, even if the intersection between the two is most likely not empty. 
Most detection methodologies are based on the density of partial sub-graphs evaluated by the incoming and outgoing degrees of the nodes. 
This approach is based on the fact that social networks have a relatively well-known and relatively stable dynamic over time (due to the social stability of human behaviour).

A sock puppet master, on the other hand, is completely free to play with and adjust the parameters for creating and animating community $\mathcal{C}$, flooding it into existing target networks. He can quantitatively and qualitatively create, remove, modify as many links as he wishes between real individuals/profiles in the network and sock puppet accounts. This will take more or less time (the more the better to remain under the radar), but it is the initial power-up phase of $\mathcal{C}$. The sock puppet mater will thus be able to ``drown'' his community in the target social networks and thwart known detection strategies. Furthermore, as the links within $\mathcal{C}$ are by nature constantly reconfigurable (dynamic communication), the ``stability'' of the community $\mathcal{C}$ is no longer assured, thus defeating the models and methods that, on the contrary, rely on the relative stability of communities.

The main approach is that without the secret key $K_0$, community identification (extraction of a high-density area for a given subset of attributes)
is impossible under these conditions. The existence or not of links between individuals is not sufficient since individuals can be in a community for for a given set of attributes without having many links with ``similar'' individuals. In absolute terms and in the context of sock puppet master activity and capabilities, brute force community detection (since $P$ can thwart known detection strategies) remains the only solution. How many communities can exist at most in a social network of order $n_v$?
\begin{itemize}
\item Without considering  the notion of attributes, this number is Bell's number (the number of partitions of a set of $n_v$ elements). It is given by the recurrence relation 
\[B_{N + 1} = \sum_{k = 0}^{n_v}\binom{n_v}{k}.B_n \quad \mbox{with} \quad B_0 = 1,\; B_1 = 1\] 
This approach is computationally intractable (for instance $B_{55} \approx 2^{177} \approx10^{53}$).
\item If we consider $n$ attributes (all those possible in a given social network, which are very substantial in number and probably constantly increasing), partitioning can be considered based on all possible subsets of attributes (once again, the mere existence or absence of links is 
not reliable with respect to (adversarial) sock puppet operations). This number is $2^n$ (total number of patterns according to the general equation~(\ref{eqmap})). Since $n$ has an order of several hundreds, it is also intractable.
\end{itemize}

Another attack consists, assuming that the community $\mathcal{C}$ is identified, of poisoning this community (inserting compatible profiles, modifying existing profiles). This attack is made impossible because, on the one hand, $P$ controls the profiles, their integrity and any modifications. Furthermore, this attack requires the knowledge of the secret key $K_0$.

Finally, assuming that the adversary has successfully identified a community $\mathcal{C}$ and the current sub-community $\mathcal{S}$, he must retrieve the current link order. For that purpose, he must break the random generator (thus recovering only the value $H^2(\mathcal{N}, K_0)$).
\subsection{Cryptanalysis Part}
Here we assume that the attacker has access to the ciphertext $C$. The cryptographic security is two-fold:
\begin{itemize}
\item The encryption algorithm (highly non linear combiner of non-linear feedback shift registers) has proven its resistance against known attack (correlation attacks, algebraic attacks).
\item The encryption algorithm enforces non trivial deniable encryption which is resistant against attacks identified with respect to the established threat model to date.
\end{itemize}
The reader can refer to~\cite{filiol44con} for more details on the encryption algorithm. 
\section{Conclusion and Future Work} \label{conc}
The CCC technology we have presented is a new approach to secure encrypted information covert communication that no longer relies on a single channel. Although further collaborative security analysis is still needed, we are convinced of the potential of this approach.

The work in progress aims to add new security primitives and new features, including:
\begin{itemize}
\item Use of weighting on nodes and edges/arcs. The aim is to increase the encoding performances with smaller sock puppet communities and larger ciphertexts.
\item Use of existing real personas/profiles within a social network (in whole or in part). Error correction techniques will be used for this purpose (real profiles can be modelled as noise).
\item Increase the cluelessness of the CCC protocol by using environmental data~\cite{riordan1998}. The idea is to make the API completely hermetic to any analysis.
\item Identify and develop additional use cases and possibly commercial applications.
\item Assess cryptographic security formally.
\end{itemize}
\begin{flushright}
	\textbf{S. D. G.}
\end{flushright}	

\end{document}